# Yellow-light Negative-index Metamaterials

Shumin Xiao, Uday K. Chettiar, Alexander V. Kildishev, Vladimir P. Drachev, and Vladimir M. Shalaev,[a]

*School of Electrical and Computer Engineering and Birck Nanotechnology Center, Purdue University, West Lafayette, Indiana 47907, USA*

[a] Electronic mail: shalaev@purdue.edu

**Abstract:**

A well-established, silver fishnet design has been further miniaturized to function as a negative-index material at the shortest wavelength to date. By studying the transmittance, reflectance, and corresponding numerical simulations of the sample, we report in this paper a negative refractive index of *-0.25* at the yellow-light wavelength of 580 nm.

Negative index metamaterials (NIMs) are artificially tailored composites of noble metals (such as silver or gold) and dielectrics [1-2] (for review see [3-4]). This kind of material has attracted a significant amount of research attention since 2000, following the prediction of superlensing using NIMs by John Pendry [5]. A near-field metamaterial superlens was successfully demonstrated soon thereafter [6]. Recently, NIM research has been further inspired by the possibility of optical cloaking [7-8]. These applications have the most to offer in the visible range, necessitating the design of NIMs that operate at less than 700 nm. Progress in visible NIMs, which began at 1.5um, has most recently been pushed to 715 nm [9]; for applications in the visible and "near-visible" ranges, recent progress in realizing negative indices in bulk metamaterials is especially important [10, 11]. Although surface plasmon-polariton (SPPs) negative-index behavior has been demonstrated at shorter wavelengths in the blue-green range, this behavior is confined in two-dimensional waveguides [12]. Unfortunately, wavelengths for three-dimensional metamaterials are still too long for superlens biological sensing and imaging applications, especially since shorter excitation wavelengths will produce images with higher optical resolution. Therefore, pushing the operational wavelength of negative refractive index materials even further into the yellow and blue range is a critical step in order to realize ever-more exciting metamaterial applications.

Since the permittivity of a noble metal is negative in the optical wavelength range, introducing a non-trivial permeability by using a magnetic resonance makes it possible to achieve a negative refractive index. Such a magnetic response can be realized in practice via the excitation of the anti-symmetric mode between two coupled metallic nanostrips. The frequency of such mode is strongly related to the size and thickness of nano-scale structures supporting the response. We therefore find that the key limitations in obtaining a magnetic resonance at shorter optical wavelengths are actually the limitations in modern fabrication techniques. For example, fishnet-based NIMs, created using e-beam lithography, are limited in their minimum achievable dimensions because the e-beam resist will collapse due to forward and backward electron scattering during electron-beam writing. Therefore, greater care is needed in both the design and fabrication of samples near the e-beam resolution limit. In this paper, we solve this problem by optimizing the design and fabrication process on a reduced unit cell in a fishnet designed for NIM structures. By carefully studying the transmittance, reflectance, and corresponding numerical simulations, we have found that the sample behaves as a NIM with the real part of its refractive index n′ = −0.25 at the yellow-light wavelength of 580 nm and a figure of merit ($FOM = n'/n''$) of *0.3*.

The fishnet design was used here because it has generally been shown to be a robust and scalable design at visible wavelengths [1, 9, 13], including the demonstration of a negative refractive index at 715 nm [9]. The structure and geometry of the fishnet are shown in Fig. 1(a), with the incident light polarization also shown for reference. The fishnet structure has periodicity in two directions – along the electric and magnetic field vector directions shown in Fig. 1(b). Before fabrication, the parameters of the fishnet design were optimized for yellow light using numerical simulations through a commercial finite-element package (COMSOL Multiphysics). These simulations give the optimum parameters for the highest *FOM* for a design with unit cell sizes of around 220 nm for both periodicities and a nanostrip bottom width of about 140 nm. As a result, the final structure has a high metal coverage ratio, or pattern density, of ~70 %.

The sample was fabricated using electron-beam lithography on poly (methyl-methacrylate) photo-resist (PMMA) followed by electron-beam evaporation and lift-off. The structure was fabricated on a glass substrate coated with a 15-nm-thick indium–tin–oxide (ITO) layer. Usually, fabrication of such a high-density fishnet is very challenging because of the forward and backward scattering of electrons within the resist, which can destroy the designed structures and cause the resist to collapse. Additionally the surface roughness of metal films can result in a broadening of the resonance, which will greatly decrease the final *FOM* [14]. In our experiment, a very low electron-beam current and a long baking time for the PMMA is used to reduce the dose for exposure. As a result, the scattering of electrons during e-beam writing was also reduced. At the same time, a low deposition rate of 0.2 Å/s was used during the deposition of the metal and dielectric layers in order to obtain a smoother final film [14].

The fabricated structure consists of two 43-nm perforated silver layers separated by a 45-nm layer of alumina. The thickness of each silver layer is slightly larger than that of previous samples, which operated at 800 nm [10].This is because the absolute value of the real part of the permittivity of silver at shorter wavelengths will be less negative than the value at longer wavelengths; therefore a larger ratio of metal to dielectric layer thicknesses is needed. The structure represents a square lattice and it has a periodic unit cell of 220 nm. A 10-nm-thick layer of alumina was deposited above and below the structure to protect the silver from deterioration and to improve the adhesion. The fishnet's stacked layers have a trapezoidal cross section due to fabrication limitations [9]. As a result, the width of the strip near the substrate (the bottom width) is larger than the top width of the strip near the air interface. Figure 1(b) shows a top view field-emission scanning electron microscope (FESEM) image of the fabricated structure. The sets of "magnetic strips" and "electric strips" are shown in darker and lighter highlights, respectively. The primary polarization is also presented with the electric field vector of the incident light perpendicular to the set of magnetic strips. The voids in the structure tend to be more rounded than the design; this is due to the backscattering of electrons. Several smaller electric strips can also be observed in Fig. 1. These smaller widths are generated because the 5 nm electron-beam spot is not always ideally circular in shape when the beam current is quite low.

The sample was optically characterized using far-field transmittance and reflectance spectra obtained with incident light polarized as defined in Fig. 1. The details of this experimental setup are described elsewhere [15]. The results are shown as solid lines in Fig. 2. For the primary polarization, two regular dips in reflection at 570 nm and 526 nm can be observed. The first dip is caused by the magnetic resonance around 570 nm, which comes from the coupling of the upper and lower layers in the magnetic strips. The second dip at 526 nm is formed by the electric resonance from the electric strips. Note that, since the electric resonance occurs at 526 nm, the electric strips behave simply like a dilute metal and provide a background negative permittivity at the longer wavelength of 570 nm. At the wavelengths of interest, even in this optimized design the magnetic resonance is still not sufficiently strong to obtain a negative permeability at 570 nm. Thus the magnetic resonance is shallow and narrow. Fortunately, the permittivity will be very negative at 570 nm under the low loss condition ($\varepsilon''$ and $\mu''$ are relatively small). As a result, the single-negative NIM (SNNIM) requirement of $\varepsilon'\mu'' + \mu'\varepsilon'' \leq 0$ is fulfilled with $\varepsilon' < 0$ and $\mu' > 0$ near the magnetic resonance [9].

In order to confirm our results and retrieve the index of refraction, numerical simulations were performed for this sample. In the simulation we used the parameters and dimensions of the fabricated sample, and the permittivity of silver was obtained from experimental data with the exception that the electron collision rate was modeled to be three times larger than that of bulk silver at the plasmon resonances [14, 16]. These model parameters correspond to the increase in electron damping due to imperfections in the silver, including grain boundaries and size effects, and due to the geometrical effects of surface roughness on the magnetic strips [14]. The simulation results are shown as dashed lines in Fig. 2. There is excellent agreement between the simulation and experimental results, including a sharp magnetic resonance near 570 nm and a broad electrical resonance near 526 nm. There is, however, a deviation in the wavelength region shorter than 500 nm. The mismatch in this shorter-wavelength range is caused by the different (than in the used model) loss of sliver at such wavelengths, which are outside of the range with negative-index behavior.

The good agreement between the spectroscopic measurements and the numerical simulation over the wide wavelength range of interest in Fig. 2 indicates the validity of our numerical model. Therefore, the effective refractive index can be calculated using the numerical results and a standard retrieval procedure [17-18]. These results are shown in Fig. 3. In Fig. 3(a), a negative refractive index in the range between 567 nm and 602 nm is observed. The lowest *n'* occurs at 580 nm where *n'* = − *0.25*. Figure 3 (b) shows the real parts of permeability and permittivity for the structure. The real part of permeability is positive throughout the entire wavelength range, while the permittivity decays towards the longer wavelengths from *–0.23* to *–0.81* in the SNNIM band from 567 nm to 602 nm. Even though the magnetic resonance is weak and the real part of the permeability is positive, a clear magnetic resonance caused by the antisymmetric mode is still present. The minimum permeability value ($\mu' \approx 0.67$) is obtained at a wavelength of 570 nm along with $\varepsilon' \approx -0.39$. This results in SNNIM behavior with $n' \approx -0.18$. The *FOM* is set to zero if $n' > 0$, and the maximum *FOM* is achieved at 580 nm (*FOM* ~ *0.3*). At 570 nm, however, the *FOM* is ~ *0.27*. The refractive index at this point is $n = -0.25 + 0.82i$, with $\varepsilon' \approx -0.47$ and $\mu' \approx 0.98$.

We also performed optical characterization and numerical simulations for the secondary polarization of incident light, in which the electric field is in plane and parallel to the magnetic strips (see Fig. 1). From the far-field spectra, the magnetic peak becomes broadened and relatively weaker. These effects can be explained by a difference in the width of the upper silver layers for electric and magnetic strips, which can be seen in Fig. 1(b). From the FESEM image, we found that there is greater variation in the width of the electric strips than in the magnetic strips. The width of the electric strips at some positions is much smaller than that of magnetic strips. Such anisotropy of the structure is likely due to the anisotropic shape of the electron beam during the e-beam lithography process. The magnetic resonance is very sensitive to the geometric parameters and to the shape of the strips. Therefore, the variation in width observed for the electric strips leads to a detuning of the resonant frequency, and as a result, the magnetic resonance becomes extremely broad and relative weak with an increase in the effective permeability, which is also positive for the secondary polarization. Hence, the requirement for SNNIM operation $\varepsilon'\mu'' + \mu'\varepsilon'' < 0$ cannot be met and the effective refractive index is positive for the secondary polarization.

In conclusion, a SNNIM sample with a fishnet structure has been fabricated and optically characterized. Negative index of refraction is for the first time achieved around 580 nm, in a 35-nm-wide wavelength band. The highest FOM of $\sim 0.3$ is observed at 580 nm along with *n'=-0.25*. Numerical simulations were also performed and are found to be in good agreement with our optical measurements in a wide wavelength range. To our knowledge, this is the first report of a yellow-light negative-refractive-index demonstration. We believe that such visible light metamaterials will find use in exciting applications in the near future.

**Acknowledgement**

This work was supported in part by ARO-MURI award 50342-PH-MUR and, in part, by NSF PREM grant # DMR 0611430.This work was supported in part by ARO-MURI award 50342-PH-MUR and, in part, by NSF PREM grant # DMR 0611430.

**Figure 1**

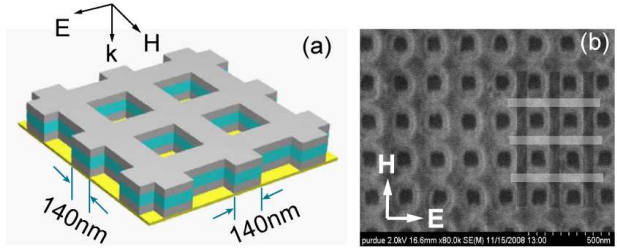

Fig. 1.

(a) Schematic of a metamaterial consisting of a fishnet structure. The geometric parameters are 220-nm periodicity along two directions, and nanostrip bottom widths are 140 nm in both directions. The layers and their thicknesses within the sample are (top to bottom): $Al_2O_3$ (10 nm) - $Ag$ (43 nm) - $Al_2O_3$ (45 nm) - $Ag$ (43 nm) - $Al_2O_3$ (10 nm) - Substrate. (b) A top view from a representative FESEM image of the structure. The primary polarization is shown in the panel and is defined as the electric field lying in the plane of the sample and perpendicular to the magnetic strips.

**Figure 2**

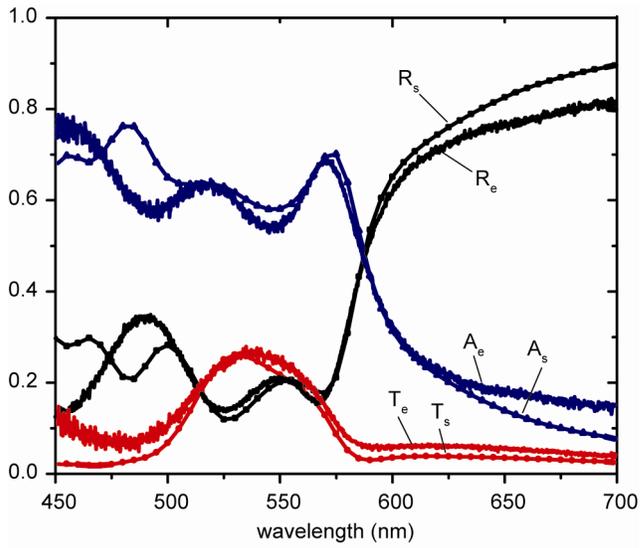

Fig. 2.

A comparison of the experimental far-field transmission ($T_e$), reflection ($R_e$) and absorption ($A_e$) spectra of the sample along with simulated results ($T_s$, $R_s$, and $A_s$) at the primary linear polarization as shown in Fig. 1(b).

**Figure 3**

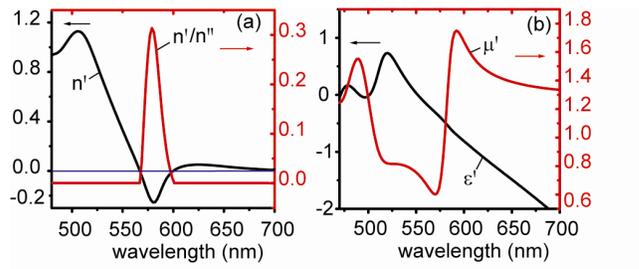

Fig. 3.

(a) Real part of the effective refractive index and FOM; the FOM is set to zero if *n>0*.

(b) Real part of the effective permeability $\mu'$ and permittivity $\varepsilon'$.